\documentclass[a4paper, 12pt, oneside]{article}
\usepackage[cp1251]{inputenc}
\usepackage[english, russian]{babel}
\usepackage{ graphics,amsfonts, amsmath,bm,amssymb, array}
\usepackage[dvips]{graphicx}
\renewcommand{\Re}{\mathop{\rm Re\,}}
\renewcommand{\Im}{\mathop{\rm Im\,}}

\makeatother {

\begin{document}
\thispagestyle{empty} \large
\renewcommand{\abstractname}{}
\renewcommand{\abstractname}{Abstract }
\renewcommand{\refname}{\begin{center} REFERENCES\end{center}}

 \begin{center}
\bf Transverse electric conductivity and dielectric function in quantum
non-degenerate collisional plasma by Mermin approach
\end{center}\medskip
\begin{center}
\bf A. V. Latyshev\footnote{$avlatyshev@mail.ru$} and
A. A. Yushkanov\footnote{$yushkanov@inbox.ru$}
\end{center}\medskip

\begin{center}
{\it Faculty of Physics and Mathematics,\\ Moscow State Regional
University, 105005,\\ Moscow, Radio str., 10--A}
\end{center}\medskip

\begin{abstract}
Formulas for transverse conductance in
quantum non-degenerate collisional plasma are deduced. The kinetic
equation in momentum space in the relaxation
approach is used. It is shown, that at $\hbar\to 0$ the derived formula
transfers to the classical one.
It is shown also, that when values of dimensionless wave numbers are small, 
the conductance formula
transfers in the known formula for classical non-degenerate plasmas.


\end{abstract}

\begin{center}
{\bf 1. Введение}
\end{center}
В хорошо известной работе Мермина \cite {Mermin} на основе анализа неравновесной
матрицы плотности в $\tau $ -- приближении было получено выражение для
продольной диэлектрической проницаемости квантовой столкновитешльной плазмы.

Ранее в работе Линдхарда \cite{Lin} были получены выражения для продольной и
поперечной диэлектрической проницаемости квантовой бесстоллкновительной плазмы.
Затем Кливер и Фукс показали \cite{Kliewer}, что прямое обобщение формул
Линдхарда на случай столкновительной плазмы (путем замены
$\omega\to \omega+i/\tau$) некорректно.
Этот недостаток для продольной диэлектрической проницаемости
был преодолен в работе Мермина \cite{Mermin}.
В тоже самое время до настоящего времени не имеется корректного выражения для
поперечной диэлектрической проницаемости для случая квантовой столкновительной
плазмы. Цель настоящей работы --- восполнить этот пробел.

Свойства электрической проводимости и диэлектрической проницаемости по
формулам, выведенным Линдхардом \cite{Lin}, подробно изучались в монографии
\cite{Dressel}. В работе \cite{Gelder} поперечная диэлектрическая проницаемость
квантовой плазмы применялась в вопросах теори скин--эффекта.
В настоящее время растет интерес к изучению различных свойств квантовой
плазмы (см, например, \cite{Anderson}--\cite{Manf2}). Особенно следует
отметить работу Дж. Манфреди \cite{Manf}, посвященную исследованию
электромагнитных свойств квантовой плазмы.

Диэлектрическая проницаемость плазмы определяется электрической
проводимостью плазмы. Поэтому сначала мы рассмотрим поперечную электрическую
проводимость квантовой столкновительной плазмы.

\begin{center}
  \bf 1. Кинетическое уравнение для матрицы плотности
\end{center}

Пусть векторный потенциал электромагнитного поля является гармоническим, т.е.
изменяется как
$
{\bf A}={\bf A}({\bf r})\exp(-i \omega t).
$
Мы рассматриваем поперечную проводимость. Поэтому выполняется следующее
соотношение
$
{\bf \rm div\bf  A}(\mathbf{r},t)=0.
$
Связь между векторным потенциалом и напряженностью электрического поля
дается следующим выражением
$$
{\bf A}({\bf q})=-\dfrac{ic}{\omega}\;{\bf E}({\bf q}).
$$

Равновесная матрица плотности имеет следующий вид
$$
{\tilde \rho}=\dfrac{1}{\exp(H/(k_BT)-\alpha)+1}.
\eqno{(1.1)}
$$

Здесь $T$ -- температура плазмы, $k_B$ --
постоянная Больцмана, $\mu$ -- химический потенциал плазмы, $H$ -- гамильтониан,
$\alpha=\mu/k_BT$ -- приведенный (безразмерный) химический
потенциал плазмы.

В линейном приближении гамильтониан имеет следующий вид
$$
H=\dfrac{({\bf p}-({e}/{c}){\bf A})^2}{2m}=
\dfrac{{\bf p}^2}{2m}-\dfrac{e}{2mc}({\bf p}{\bf A}+{\bf A} {\bf
p}).
$$

Здесь $\mathbf{p}$ -- оператор импульса, $\mathbf{p}=-i\hbar \nabla$,
$e, m$ -- заряд и масса электрона, $c$ -- скорость света.

Следовательно, мы можем представить этот гамильтониан в виде суммы
двух операторов $H=H_0+H_1$,
где
$$
H_0=\dfrac{{\bf p}^2}{2m},
\qquad H_1=-\dfrac{e}{2mc}({\bf p}{\bf A}+{\bf A} {\bf p}).
$$

Возьмем кинетическое уравнение для матрицы плотности в $\tau$ -- приближении
$$
i\hbar \dfrac{\partial \rho}{\partial t}=[H,\rho]+ i\hbar
\dfrac{\tilde\rho-\rho}{\tau}.
\eqno{(1.2)}
$$

Здесь $\nu=1/\tau$ -- эффективная частота столкновений частиц плазмы,
$\tau$ -- характерное время между двумя последовательными столкновениями,
$\hbar$ -- постоянная Планка, $[H,\rho]=H\rho-\rho H$ -- коммутатор,
$\mathbf{\tilde{\rho}}$ -- равновесная матрица плотности.

В линейном приближении по внешнему полю мы ищем матрицу плотности в виде
$$
{\rho}={\tilde \rho}^{(0)}+{\rho}^{(1)}.
\eqno{(1.3)}
$$

Здесь ${\rho}^{(1)}$ -- поправка (возмущение) к равновесной
матрице плотности, обусловленная наличием электромагнитного поля,
$\tilde{\rho}^{(0)}$ -- равновесная матрица плотности,
отвечающая "равновесному"\, оператору Гамильтона $H_0$.

Представим равновесную матрицу плотности $\tilde{\rho}$ (см. (1.1))
в следующем виде
$$
\tilde{\rho}=\tilde{\rho}^{(0)}+\tilde{\rho}^{(1)}.
\eqno{(1.4)}
$$

Рассмотрим коммутатор $[H, \tilde{\rho}]$. В линейном приближении
этот коммутатор равен
$$
[H, {\tilde \rho}\,]=[H_0, {\tilde \rho}^{(1)}]+[H_1, {\tilde
\rho}^{(0)}]
\eqno{(1.5)}
$$
и
$$
[H, {\tilde \rho}\,]=0.
\eqno{(1.6)}
$$

Для коммутаторов из правой части равенства (1.5) мы находим
$$
\langle\mathbf{k}_1|[H_0, \tilde{\rho}^{(1)}]|\mathbf{k}_2\rangle=
\big(E_{\mathbf{k}_1}-E_{\mathbf{k}_2}\big)\tilde{\rho}^{(1)}
(\mathbf{k}_1-\mathbf{k}_2),
\eqno{(1.7)}
$$
и
$$
\langle\mathbf{k}_1|[H_1, \tilde{\rho}^{(0)}]|\mathbf{k}_2\rangle=
\Big[f_F(\mathbf{k}_2)-f_F(\mathbf{k}_1)\Big]\langle\mathbf{k}_1|H_1|
\mathbf{k}_2\rangle=
$$
$$
=\dfrac{e\hbar}{2mc}\Big[f_F(\mathbf{k}_2)-f_F(\mathbf{k}_1)\Big]
(\mathbf{k}_1+\mathbf{k}_2)\mathbf{A}(\mathbf{k}_1-\mathbf{k}_2).
\eqno{(1.8)}
$$
Здесь
$$
f_F(\mathbf{k})=\dfrac{1}{1+\exp ( E_{\mathbf{k}}/k_BT-\alpha)},
$$
где
$$
E_{\mathbf{k}}=\dfrac{\hbar^2\mathbf{k}^2}{2m}, \qquad
\mathbf{p}=\hbar \mathbf{k}.
$$

Напомним, что вектор $|\mathbf{k}\rangle$ является собственным вектором
оператора $H$;
$$
H|\mathbf{k}\rangle =E_{\mathbf{k}} |\mathbf{k}\rangle, \qquad
\langle \mathbf{k}|H=E_{\mathbf{k}}\langle \mathbf{k}|,
$$
и оператора $\mathbf{p}$:
$$
\mathbf{p}|\mathbf{k}\rangle =\hbar \mathbf{k} |\mathbf{k}\rangle, \qquad
\langle \mathbf{k}|\mathbf{p}=\hbar \mathbf{k}\langle \mathbf{k}|.
$$

Напомним также, что для оператора $L$ выполняется соотношение:
$$
\langle \mathbf{k}_1|L|\mathbf{k}_2\rangle =\dfrac{1}{(2\pi)^3}\int
\exp(-i\mathbf{k}_1\mathbf{r})L\exp(i\mathbf{k}_2\mathbf{r})d\mathbf{r},
$$
а для функции $\varphi(\mathbf{r})$ -- такое соотношение
$$
\langle \mathbf{k}_1|\varphi|\mathbf{k}_2\rangle =\dfrac{1}{(2\pi)^3}\int
\exp[-i(\mathbf{k}_1-\mathbf{k}_2)\mathbf{r})\varphi(\mathbf{r})d\mathbf{r}=
\varphi(\mathbf{r}_1-\mathbf{r}_2).
$$

Из соотношений (1.4)--(1.8) вытекает, что
$$
\tilde{\rho}^{(1)}(\mathbf{k}_1-\mathbf{k}_2)=
-\dfrac{e\hbar}{2mc}\dfrac{f_F(\mathbf{k}_1)-
f_F(\mathbf{k}_2)}{E_{\mathbf{k}_1}-E_{\mathbf{k}_2}}
(\mathbf{k}_1+\mathbf{k}_2)\mathbf{A}(\mathbf{k}_1-\mathbf{k}_2).
\eqno{(1.9)}
$$

С помощью соотношений (1.3)--(1.5) мы линеаризуем кинетическое уравнение
(1.2).
Получаем следующее уравнение
$$
i\hbar \dfrac{\partial \rho^{(1)}}{\partial t}=[H_0,
\rho^{(1)}]+[H_1, \tilde{\rho}^{(0)}]+i\hbar\dfrac{\tilde{\rho}^{(1)}-
\rho^{(1)}}{\tau}.
\eqno{(1.10)}
$$

Заметим, что возмущение $\rho^{(1)}\sim \exp(-i\omega t)$,
Тогда уравнение (1.10) принимает следующий вид
$$
(\hbar \omega+i\hbar \nu)\rho^{(1)}=[H_0, \rho^{(1)}]+
[H_1, \tilde{\rho}^{(0)}]+i\hbar \nu \tilde{\rho}^{(1)}.
$$
Отсюда следует, что
$$
(\hbar \omega+i\hbar \nu)\langle\mathbf{k}_1|\rho^{(1)}|\mathbf{k}_2\rangle=
\langle\mathbf{k}_1|[H_0,\rho^{(1)}]|\mathbf{k}_2\rangle+
$$
$$
+\langle\mathbf{k}_1|[H_1,\tilde{\rho}^{(0)}]|\mathbf{k}_2\rangle+i\hbar \nu
\langle\mathbf{k}_1|\tilde{\rho}^{(1)}|\mathbf{k}_2\rangle.
\eqno{(1.11)}
$$

Нетрудно видеть, что $$
\langle\mathbf{k}_1|[H_0,\rho^{(1)}]|\mathbf{k}_2\rangle=(E_{\mathbf{k}_1}-
E_{\mathbf{k}_2})\langle\mathbf{k}_1|\rho^{(1)}|\mathbf{k}_2\rangle=$$$$=
(E_{\mathbf{k}_1}-E_{\mathbf{k}_2})\rho^{(1)}(\mathbf{k}_1-\mathbf{k}_2).
\eqno{(1.12)}
$$

Подставим в (1.11) равенства (1.12) и (1.8). В результате будем иметь: $$
(\hbar\omega+i\hbar \nu-E_{\mathbf{k}_1}+E_{\mathbf{k}_2})
\rho^{(1)}(\mathbf{k}_1-\mathbf{k}_2)=
$$
$$
=\dfrac{e\hbar}{2mc}[f_F(\mathbf{k}_1)-f_F(\mathbf{k}_2)]
(\mathbf{k}_1+\mathbf{k}_2)\mathbf{A}(\mathbf{k}_1-\mathbf{k}_2)+
i\hbar \nu \tilde{\rho}^{(1)}(\mathbf{k}_1-\mathbf{k}_2).
\eqno{(1.13)}
$$

Преобразуем теперь уравнение (1.13) с помощью равенства (1.9) к виду
$$
(\hbar\omega+i\hbar \nu-E_{\mathbf{k}_1}+E_{\mathbf{k}_2})
\rho^{(1)}(\mathbf{k}_1-\mathbf{k}_2)=
$$
$$
=\dfrac{e\hbar}{2mc}\dfrac{[f_F(\mathbf{k}_1)-f_F(\mathbf{k}_2)]
(E_{\mathbf{k}_1}-E_{\mathbf{k}_2}-i\hbar \nu)}{E_{\mathbf{k}_1}-
E_{\mathbf{k}_2}}(\mathbf{k}_1+\mathbf{k}_2)\mathbf{A}(\mathbf{k}_1-
\mathbf{k}_2),
$$
из которого находим
$$
\rho^{(1)}(\mathbf{k}_1-\mathbf{k}_2)=
$$\hspace{0.1cm}
$$=
\dfrac{e\hbar}{2mc}
\dfrac{[f_F(\mathbf{k}_1)-f_F(\mathbf{k}_2)]
(E_{\mathbf{k}_1}-E_{\mathbf{k}_2}-i\hbar \nu)}
{(E_{\mathbf{k}_1}-E_{\mathbf{k}_2})(\hbar\omega+i\hbar \nu-
E_{\mathbf{k}_1}+E_{\mathbf{k}_2})}
(\mathbf{k}_1+\mathbf{k}_2)\mathbf{A}(\mathbf{k}_1-\mathbf{k}_2).
\eqno{(1.14)}
$$\hspace{0.2cm}

В уравнении (1.14) мы положим $\mathbf{k}_1=\mathbf{k}$,
$\mathbf{k}_2=\mathbf{k}-\mathbf{q}$. Тогда
$$
\langle\mathbf{k}_1|\rho^{(1)}|\mathbf{k}_2\rangle=
\langle\mathbf{k}|\rho^{(1)}|\mathbf{k}-
\mathbf{q}\rangle= \rho^{(1)}(\mathbf{q})=
$$ \hspace{0.2cm}
$$
=-\dfrac{e\hbar}{mc}
\dfrac{[f_F(\mathbf{k})-f_F(\mathbf{k}-\mathbf{q})]
(E_{\mathbf{k}}-E_{\mathbf{k}-\mathbf{q}}-i\hbar \nu)}
{(E_{\mathbf{k}}-E_{\mathbf{k}-\mathbf{q}})(
E_{\mathbf{k}}-E_{\mathbf{k}-\mathbf{q}}-\hbar\omega-i\hbar\nu)}
\mathbf{k}\mathbf{A}(\mathbf{q}).
\eqno{(1.15)}
$$\hspace{0.3cm}

\begin{center}
  \bf 2. Плотность тока
\end{center}

Плотность тока ${\bf j}({\bf q})$ определяется как
$$
{\bf j}({\bf q},\omega)=e\int \dfrac{d{\bf k}}{8\pi^3m}\left\langle{\bf k}+
\frac{\mathbf{q}}{2}\left|({\bf p}-\frac{e}{c}{\bf A})
\rho+\rho ({\bf p}-\frac{e}{c}{\bf A} \big)\right|{\bf k}-
\frac{\mathbf{q}}{2}\right\rangle.
\eqno{(2.1)}
$$
После подстановки (1.3) в интеграл из (2.1), мы имеем
$$
\left\langle{\bf k}+
\frac{\mathbf{q}}{2}\left|({\bf p}-\frac{e}{c}{\bf A})
\rho+\rho ({\bf p}-\frac{e}{c}{\bf A} \big)\right|{\bf k}-
\frac{\mathbf{q}}{2}\right\rangle=$$$$=
\left\langle{\bf k}+
\frac{\mathbf{q}}{2}\left|{\bf p}\rho^{(1)}+\rho^{(1)}\mathbf{p}-\frac{e}{c}
({\bf A}\tilde{\rho}^{(0)}+\tilde{\rho}^{(0)}\mathbf{A})\right|{\bf k}-
\frac{\mathbf{q}}{2}\right\rangle.
$$
Нетрудно показать, что
$$
\left\langle{\bf k}+\dfrac{\mathbf{q}}{2}\left|{\bf p}\rho^{(1)}+
\rho^{(1)}\mathbf{p}\right|{\bf k}-
\dfrac{\mathbf{q}}{2}\right\rangle
=2\hbar\mathbf{k}\rho^{(0)}(\mathbf{q}),
$$
$$
\left\langle{\bf k}+
\dfrac{\mathbf{q}}{2}\left|{\bf A}\tilde{\rho}^{(0)}+
\tilde{\rho}^{(0)}\mathbf{A}\right|{\bf k}-\dfrac{\mathbf{q}}{2}\right\rangle=
\mathbf{A}(\mathbf{q})\Big[\tilde{\rho}^{(0)}(\mathbf{k}+
\dfrac{\mathbf{q}}{2})+\tilde{\rho}^{(0)}(\mathbf{k}-
\dfrac{\mathbf{q}}{2})\Big].
$$

Следовательно, выражение для плотности тока имеет следующий вид
$$
\mathbf{j}(\mathbf{q},\omega,\nu)=-\dfrac{e^2}{mc}\mathbf{A}(\mathbf{q})
\int \dfrac{d\mathbf{k}}{8\pi^3}\tilde{\rho}^{(0)}(\mathbf{k}+
\dfrac{\mathbf{q}}{2})-\dfrac{e^2}{mc}\mathbf{A}(\mathbf{q})
\int \dfrac{d\mathbf{k}}{8\pi^3}\tilde{\rho}^{(0)}(\mathbf{k}-
\dfrac{\mathbf{q}}{2})+
$$
$$
+e\hbar\int\dfrac{d\mathbf{k}}{4\pi^3m}
\left\langle\mathbf{k}+\dfrac{\mathbf{q}}{2}
\left|\rho^{(1)}\right|\mathbf{k}-\dfrac{\mathbf{q}}{2}\right\rangle.
$$

Первые два члена в этом выражении равны друг другу
$$
\int \dfrac{d\mathbf{k}}{8\pi^3}\tilde{\rho}^{(0)}(\mathbf{k}+
\dfrac{\mathbf{q}}{2})=
\int \dfrac{d\mathbf{k}}{8\pi^3}\tilde{\rho}^{(0)}(\mathbf{k}-
\dfrac{\mathbf{q}}{2})=\dfrac{N}{2},
$$
где $N$ -- числовая плотность (концентрация) плазмы.

Следовательно, плотность тока равна
$$
{\bf j}({\bf q},\omega,\nu)=-\frac{e^2N}{mc}{\bf A}({\bf q})+
e\hbar\int \dfrac{d{\bf k}}{4\pi^3m}{\bf k}
\left\langle\mathbf{k}+\dfrac{\mathbf{q}}{2}\left|\rho^{(1)}\right|{\bf k}-
\frac{\mathbf{q}}{2}\right\rangle.
\eqno{(2.2)}
$$

Первое слагаемое в (2.2) есть не что иное, как калибровочная плотность тока.

С помощью очевидной замены переменных в интеграле из (2.2) выражение (2.2)
можно преобразовать к виду
$$
{\bf j}({\bf q},\omega,\nu)=-\frac{e^2N}{mc}{\bf A}({\bf q})+
e\hbar\int \dfrac{d{\bf k}}{4\pi^3m}{\bf k}
\left\langle\mathbf{k}\left|\rho^{(1)}\right|{\bf k}-{\bf q}\right\rangle.
\eqno{(2.3)}
$$

В соотношении (2.3) подынтегральное выражение дается равенством (1.12).
Подставляя (1.15) в (2.3), получаем следующее выражение для плотности тока
$$
{\bf j}({\bf q},\omega,\nu)=-\frac{e^2N}{mc}{\bf A}({\bf q})-
$$
$$
-\dfrac{e^2\hbar^2}{m^2c}\int \dfrac{\mathbf{k}d\mathbf{k}}{4\pi^3}[\mathbf{k}
\mathbf{A(q)}]\dfrac{[f_F(\mathbf{k})-f_F(\mathbf{k-q})]
(E_{\mathbf{k}}-E_{\mathbf{k-q}}-i\hbar \nu)}{(E_{\mathbf{k}}-
E_{\mathbf{k-q}})(E_{\mathbf{k}}-E_{\mathbf{k-q}}-\hbar \omega -
i\hbar \nu)}.
\eqno{(2.4)}
$$

Обозначим далее
$$
\Xi(\mathbf{k,q})=\dfrac{E_{\mathbf{k}}-E_{\mathbf{k-q}}-i\hbar \nu}
{(E_{\mathbf{k}}-
E_{\mathbf{k-q}})[E_{\mathbf{k}}-E_{\mathbf{k-q}}-\hbar (\omega +i\nu)]}.
$$

С учетом этого обозначения формулу (2.4) для плотности тока можно переписать
короче:
$$
{\bf j}({\bf q},\omega,\nu)=-\frac{e^2N}{mc}{\bf A}({\bf q})- $$$$-
\dfrac{e^2\hbar^2}{m^2c}\int \dfrac{\mathbf{k}d\mathbf{k}}{4\pi^3}[\mathbf{k}
\mathbf{A(q)}]\;\Xi(\mathbf{k,q})[f_F(\mathbf{k})-f_F(\mathbf{k-q})].
$$

Направим ось $x$ вдоль вектора ${\bf q}$, а ось $y$ вдоль
вектора ${\bf A}$. Тогда предыдущее векторное выражение (2.4) может быть
переписано в виде трех скалярных
$$
{ j}_y({\bf q},\omega,\nu)=-\frac{e^2N}{mc}{ A}({\bf q})-
\dfrac{e^2\hbar^2 A({\bf q})}{m^2c}\int \dfrac{d{\bf k}}{4\pi^3}{ k}_y^2
\;\Xi(\mathbf{k,q})[f_F(\mathbf{k})-f_F(\mathbf{k-q})] $$
и
$$
{ j}_x({\bf q},\omega,\nu)={ j}_z({\bf q},\omega,\nu)=0.
$$

Очевидно, что
$$
\int \dfrac{d{\bf k}}{4\pi^3}{ k}_y^2\;\Xi(\mathbf{k,q})
[f_F(\mathbf{k})-f_F(\mathbf{k-q})] =\int \dfrac{d{\bf k}}{4\pi^3}{ k}_z^2\;\Xi(\mathbf{k,q})
[f_F(\mathbf{k})-f_F(\mathbf{k-q})]. $$
Следовательно
$$
\int \dfrac{d{\bf k}}{4\pi^3}{ k}_y^2\;\Xi(\mathbf{k,q})
[f_F(\mathbf{k})-f_F(\mathbf{k-q})]=
$$
$$
\dfrac{1}{2}\int \dfrac{d{\bf k}}{4\pi^3}({ k}_y^2+{
k}_z^2)\;\Xi(\mathbf{k,q})[f_F(\mathbf{k})-f_F(\mathbf{k-q})]= $$$$=
\dfrac{1}{2}\int \dfrac{d{\bf k}}{4\pi^3}({\bf k}^2-{
k}_x^2)\;\Xi(\mathbf{k,q})[f_F(\mathbf{k})-f_F(\mathbf{k-q})].
$$
Отсюда мы заключаем, что выражение для плотности тока можно представить в
следующей инвариантной форме
$$
{\bf j}({\bf q},\omega,\nu)=-\frac{Ne^2}{mc}{\bf A}({\bf q})-\hspace{6cm}
$$
$$
-\dfrac{e^2\hbar^2}{8\pi^3m^2c}{\bf A}({\bf q})\int d{\bf k}
\Big[{\bf k}^2-\Big(\dfrac{{\bf k}{\bf q}}{q}\Big)^2\Big]\;\Xi(\mathbf{k,q})
[f_F(\mathbf{k})-f_F(\mathbf{k-q})],
\eqno{(2.5)} $$
или, учитывая разложение на элементарные дроби,  (2.5) запишем в виде
$$
\dfrac{E_{\mathbf{k}}-E_{\mathbf{k-q}}-i\hbar \nu}{(E_{\mathbf{k}}-
E_{\mathbf{k-q}})(E_{\mathbf{k}}-E_{\mathbf{k-q}}-\hbar \omega -
i\hbar \nu)}=
$$
$$
=\dfrac{1}{E_{\mathbf{k}}-E_{\mathbf{k-q}}}+\dfrac{\hbar \omega}
{E_{\mathbf{k}}-E_{\mathbf{k-q}}-\hbar(\omega+i \nu)},
$$
получаем
$$
{\bf j}({\bf q},\omega,\nu)=-\frac{Ne^2}{mc}{\bf A}({\bf q})-
$$
$$
-\dfrac{e^2\hbar^2}{8\pi^3m^2c}{\bf A}({\bf q})\int\mathbf{k}_\perp^2 d{\bf k}
\dfrac{f_F({\bf k})-f_F({\bf k-q})}{E_{{\bf k}}-E_{{\bf k-q}}}-
$$
$$
-\dfrac{e^2\hbar^3\omega}{8\pi^3m^2c}{\bf A}({\bf q})\int \mathbf{k}_\perp^2
d{\bf k}
\dfrac{f_F({\bf k})-f_F({\bf k-q})}{(E_{{\bf k}}-E_{{\bf k-q}})
[ E_{{\bf k}}-E_{{\bf k-q}}-\hbar( \omega+i\nu)]}.
\eqno{(2.6)}
$$\medskip

Здесь
$$
\mathbf{k}_\perp^2= {\bf k}^2-\Big(\dfrac{{\bf k}{\bf q}}{q}\Big)^2.
$$

Первые два члена в предыдущем соотношении (2.6) не зависят от частоты
$\omega$ и определяются диссипативными свойствами материала, определяемыми
частотой столкновений $\nu$. Эти члены являются универсальными параметрами,
определяющими диамагнетизм Ландау.

\begin{center}
\bf 3. Поперечная электрическая проводимость и диэлектрическая проницаемость
\end{center}

Учитывая связь векторного потенциала с напряженностью электромагнитного поля,
а также связь плотности тока с электрическим полем, на основании предыдущих соотношений (2.5) и (2.6) получаем следующее выражение инвариантного вида
для поперечной электрической проводимости
$$
\sigma_{tr}(\mathbf{q},\omega,\nu)=
\dfrac{ie^2N}{m\omega}+\dfrac{ie^2\hbar^2}{8\pi^3m^2\omega}
\int \Xi(\mathbf{k,q})[f_F({\bf k})-f_F({\bf k-q})]
\mathbf{k}_\perp^2 d\mathbf{k}.
$$
или, выделяя статическую проводимость $\sigma_0=e^2N/m \nu$,
$$
\dfrac{\sigma_{tr}(\mathbf{q},\omega,\nu)}{\sigma_0}=
\dfrac{i \nu}{\omega}\Bigg[1+\dfrac{\hbar^2}{8\pi^3m N}
\int  \Xi(\mathbf{k,q})[f_F({\bf k})-f_F({\bf k-q})]
\mathbf{k}_\perp^2 d\mathbf{k}\Bigg].
\eqno{(3.1)}
$$

Представим соотношение (3.1) в явном виде
$$
\dfrac{\sigma_{tr}(\mathbf{q},\omega,\nu)}{\sigma_0}=
\dfrac{i \nu}{\omega}\Bigg[1+ $$$$ +
\dfrac{\hbar^2}{8\pi^3mN}\int \dfrac{[f_F(\mathbf{k})-f_F(\mathbf{k-q})]
(E_{\mathbf{k}}-E_{\mathbf{k-q}}-i\hbar \nu)}{(E_{\mathbf{k}}-
E_{\mathbf{k-q}})[E_{\mathbf{k}}-E_{\mathbf{k-q}}-\hbar (\omega +i\nu)]}
\mathbf{k}_\perp^2 d\mathbf{k}\Bigg].
\eqno{(3.2)}
$$

На основании (3.1) и (3.2) напишем выражение для диэлектрической проницаемости
$$
\varepsilon_{tr}=1-\dfrac{\omega_p^2}{\omega^2}\Bigg[1+
\dfrac{\hbar^2}{8\pi^3m N}
\int  \Xi(\mathbf{k,q})[f_F({\bf k})-f_F({\bf k-q})]
\mathbf{k}_\perp^2 d\mathbf{k}\Bigg]
\eqno{(3.3)}
$$
и
$$
\varepsilon_{tr}=1-\dfrac{\omega_p^2}{\omega^2}\Bigg[1+
 $$$$ +
\dfrac{\hbar^2}{8\pi^3mN}\int \dfrac{[f_F(\mathbf{k})-f_F(\mathbf{k-q})]
(E_{\mathbf{k}}-E_{\mathbf{k-q}}-i\hbar \nu)}{(E_{\mathbf{k}}-
E_{\mathbf{k-q}})[E_{\mathbf{k}}-E_{\mathbf{k-q}}-\hbar (\omega +i\nu)]} \mathbf{k}_\perp^2 d\mathbf{k}\Bigg].
\eqno{(3.4)}
$$

Заметим, что для невырожденной плазмы числовая плотность в равновесном
состоянии равна
$$
N=\dfrac{f_2(\alpha)}{\pi^2}k_T^3,
$$
где $k_T$ -- тепловое волновое число, $k_T=\dfrac{mv_T}{\hbar},
v_T=\dfrac{1}{\sqrt{\beta}}$, $\beta=\dfrac{m}{2k_BT}$,
$$
f_2(\alpha)=\int\limits_{0}^{\infty}x^2f_F(x)dx=\int\limits_{0}^{\infty}
\dfrac{x^2dx}{1+e^{x^2-\alpha}}=\dfrac{1}{2}\int\limits_{0}^{\infty}
\ln(1+e^{\alpha-x^2})dx.
$$

Преобразуем формулы (3.1), (3.2) и (3.3), (3.4). Вместо вектора
$\mathbf{k}$ введем
безразмерный вектор $\mathbf{P}$ равенством
$$
\mathbf{P}=\dfrac{\mathbf{k}}{k_T}.
$$
Возьмем вектор $\mathbf{q}=q(1,0,0)$. Тогда
$$
\mathbf{k}_\perp^2=\mathbf{k}^2-\Big(\dfrac{\mathbf{kq}}{q}\Big)^2=
k^2-k_x^2=k_T^2(P^2-P_x^2)=k_T^2P_\perp^2.
$$

Таким образом, формулы (3.1) и (3.2) запишутся в виде:
$$
\dfrac{\sigma_{tr}}{\sigma_0}=\dfrac{i \nu}{\omega}\Bigg[1+
\dfrac{mv_T^2}{8\pi f_2(\alpha)}\int \Xi(\mathbf{k,q})
[f_F({\bf k})-f_F({\bf k-q})]P_\perp^2d^3P\Bigg]
$$ и
$$
\dfrac{\sigma_{tr}}{\sigma_0}=\dfrac{i \nu}{\omega}\Bigg[1+
\dfrac{mv_T^2}{8\pi f_2(\alpha)}\int \dfrac{[f_F(\mathbf{k})-f_F(\mathbf{k-q})]
(E_{\mathbf{k}}-E_{\mathbf{k-q}}-i\hbar \nu)}{(E_{\mathbf{k}}-
E_{\mathbf{k-q}})[E_{\mathbf{k}}-E_{\mathbf{k-q}}-\hbar (\omega +i\nu)]}
P_\perp^2d^3P\Bigg].
\eqno{(3.5)}
$$

Точно так же преобразуются формулы (3.3) и (3.4):
$$
\varepsilon_{tr}=1-\dfrac{\omega_p^2}{\omega^2}\Bigg[1+
\dfrac{mv_T^2}{8\pi f_2(\alpha)}
\int  \Xi(\mathbf{k,q})[f_F({\bf k})-f_F({\bf k-q})]P_\perp^2d^3P\Bigg]
$$
и
$$
\varepsilon_{tr}=1-\dfrac{\omega_p^2}{\omega^2}\Bigg[1+
 $$$$ +
\dfrac{mv_T^2}{8\pi f_2(\alpha)}\int \dfrac{[f_F(\mathbf{k})-f_F(\mathbf{k-q})]
(E_{\mathbf{k}}-E_{\mathbf{k-q}}-i\hbar \nu)}{(E_{\mathbf{k}}-
E_{\mathbf{k-q}})[E_{\mathbf{k}}-E_{\mathbf{k-q}}-\hbar (\omega +i\nu)]}
P_\perp^2d^3P\Bigg].
\eqno{(3.6)}
$$

Воспользуемся разложением на элементарные дроби $$
\dfrac{E_{\mathbf{k}}-E_{\mathbf{k-q}}-i\hbar \nu}
{(E_{\mathbf{k}}-E_{\mathbf{k-q}})[E_{\mathbf{k}}-E_{\mathbf{k-q}}-\hbar
(\omega+i\nu)]}=
$$
$$
=\dfrac{i \nu}{\omega+i \nu}\dfrac{1}
{E_{\mathbf{k}}-E_{\mathbf{k-q}}}+\dfrac{\omega}{\omega+i \nu}\dfrac{1}
{E_{\mathbf{k}}-E_{\mathbf{k-q}}-\hbar (\omega+i\nu)}.
$$
С помощью этого разложения перепишем формулы (3.5) и (3.6):
$$
\dfrac{\sigma_{tr}}{\sigma_0}=\dfrac{i \nu}{\omega}\Bigg[1+
\dfrac{mv_T^2}{8\pi f_2(\alpha)}\dfrac{i \nu}{\omega+i \nu}
\int\dfrac{f_F(\mathbf{k})-f_F(\mathbf{k-q})}{E_{\mathbf{k}}-E_{\mathbf{k-q}}}
P_\perp^2d^3P+
$$
$$  +
\dfrac{mv_T^2}{8\pi f_2(\alpha)}\dfrac{\omega}{\omega+i \nu}
\int\dfrac{f_F(\mathbf{k})-f_F(\mathbf{k-q})}{E_{\mathbf{k}}-E_{\mathbf{k-q}}
-\hbar(\omega+i \nu)}P_\perp^2d^3P\Bigg]
\eqno{(3.7)}
$$
и
$$
\varepsilon_{tr}=1-\dfrac{\omega_p^2}{\omega^2}\Bigg[1+
\dfrac{mv_T^2}{8\pi f_2(\alpha)}\dfrac{i \nu}{\omega+i \nu}
\int\dfrac{f_F(\mathbf{k})-f_F(\mathbf{k-q})}{E_{\mathbf{k}}-E_{\mathbf{k-q}}}
P_\perp^2d^3P+
$$
$$  +
\dfrac{mv_T^2}{8\pi f_2(\alpha)}\dfrac{\omega}{\omega+i \nu}
\int\dfrac{f_F(\mathbf{k})-f_F(\mathbf{k-q})}{E_{\mathbf{k}}-E_{\mathbf{k-q}}
-\hbar(\omega+i \nu)}P_\perp^2d^3P \Bigg].
\eqno{(3.8)}
$$

Рассмотрим интегралы
$$
J_\nu=\dfrac{1}{2\pi}\int  \dfrac{f_F(\mathbf{k})-f_F(\mathbf{k-q})}
{E_{\mathbf{k}}-E_{\mathbf{k-q}}}P_\perp^2d^3P
$$
и
$$
J_\omega=\dfrac{1}{2\pi}\int  \dfrac{f_F(\mathbf{k})-f_F(\mathbf{k-q})}
{E_{\mathbf{k}}-E_{\mathbf{k-q}}-\hbar(\omega+i \nu)}P_\perp^2d^3P.
$$

С помощью этих интегралов формулы (3.7) и (3.8) перепишутся в симметричной
форме:
$$
\dfrac{\sigma_{tr}}{\sigma_0}=\dfrac{i \nu}{\omega}
\Big[1+\dfrac{mv_T^2}{4f_2(\alpha)}\cdot\dfrac{\omega J_\omega+i \nu J_\nu}
{\omega+i \nu}\Big]
\eqno{(3.9)}
$$
и
$$
\varepsilon_{tr}=1-\dfrac{\omega_p^2}{\omega^2}
\Big[1+\dfrac{mv_T^2}{4f_2(\alpha)}\cdot\dfrac{\omega J_\omega+i \nu J_\nu}
{\omega+i \nu}\Big].
\eqno{(3.10)}
$$

Представим $J_\nu$ в виде разности двух интегралов. Во втором интеграле сделаем
очевидную замену переменной интегрирования. В результате получаем:
$$
J_\nu=\dfrac{1}{2\pi}\int \dfrac{2E_{\mathbf{k}}-(E_{\mathbf{k-q}}-E_{\mathbf{k+q}})}
{(E_{\mathbf{k}}-E_{\mathbf{k-q}})(E_{\mathbf{k}}-E_{\mathbf{k+q}})}
f_F(\mathbf{k})P_\perp^2d^3P.
$$
Аналогично преобразуем второй интеграл
$$
J_\omega=\dfrac{1}{2\pi}\int
\dfrac{[2E_{\mathbf{k}}-(E_{\mathbf{k-q}}-E_{\mathbf{k+q}})]
f_F(\mathbf{k})P_\perp^2d^3P}
{(E_{\mathbf{k}}-E_{\mathbf{k-q}}-\hbar(\omega+i \nu))
(E_{\mathbf{k}}-E_{\mathbf{k+q}}+\hbar(\omega+i \nu))}.
$$

Займемся преобразованием интегралов $J_\nu$ и $J_\omega$.
Энергия электрона равна
$$
E_{\mathbf{k}}=\dfrac{\hbar^2\mathbf{k}^2}{2m}=\dfrac{\hbar^2k_T^2P^2}{2m}=
\dfrac{mv_T^2}{2}P^2=E_TP^2,
$$
где $E_T=\dfrac{mv_T}{2}$ -- тепловая энергия электронов.

Аналогично,
$$
E_{\mathbf{k\mp q}}=\dfrac{\hbar^2}{2m}(\mathbf{k\mp q})^2=E_T\Big(\mathbf{P}
\mp\dfrac{\mathbf{q}}{k_T}\Big)^2=E_T(\mathbf{P\mp l})^2,
$$
где $\mathbf{l}=\dfrac{\mathbf{q}}{k_T}=\dfrac{\hbar\mathbf{q}}{mv_T}$ --
безразмерное волновое число.

Разность этих величин равна:
$$
E_{\mathbf{k}}-E_{\mathbf{k-q}}=E_T(2P_xl-l^2)=lmv_T(P_x-\dfrac{l}{2}),
$$
$$
E_{\mathbf{k}}-E_{\mathbf{k+q}}=-E_T(2P_xl+l^2)=-lmv_T(P_x+\dfrac{l}{2}).
$$
Найдем сумму этих равенств:
$$
2E_{\mathbf{k}}-(E_{\mathbf{k+q}}+E_{\mathbf{k-q}})=-2l^2E_T=-l^2mv_T^2.
$$

Значит,
$$
J_\nu=\dfrac{1}{2\pi mv_T^2}\int \dfrac{f_F(P)P_\perp^2d^3P}{P_x^2-(l/2)^2}.
$$

Далее, получаем:
$$
E_{\mathbf{k}}-E_{\mathbf{k-q}}-\hbar(\omega+i \nu)=lmv_T^2(P_x-
\dfrac{l}{2})-
\hbar(\omega+i \nu)=$$
$$
=lmv_T^2\Big[P_x-\dfrac{l}{2}-\dfrac{\hbar(\omega+i \nu)}
{lmv_T^2}\Big]=lmv_T^2\Big[P_x-\dfrac{l}{2}-\dfrac{z}{l}\Big],
$$
где
$$
z=\dfrac{\omega+i \nu}{k_Tv_T}=x+iy,\qquad x=\dfrac{\omega}{k_Tv_T},\qquad
y=\dfrac{\nu}{k_Tv_T}.
$$
Точно так же
$$
E_{\mathbf{k}}-E_{\mathbf{k-q}}+\hbar(\omega+i \nu)=
lmv_T^2\Big[P_x+\dfrac{l}{2}-\dfrac{z}{l}\Big].
$$
Значит,
$$
J_\omega=\dfrac{1}{2\pi mv_T^2}
\int \dfrac{f_F(P)P_\perp^2d^3P}{(P_x-z/l)^2-(l/2)^2}.
$$

В интегралах $J_\nu$ и $J_\omega$ вычислим внутренний двойной интеграл:
$$
\int\limits_{-\infty}^{\infty}\int\limits_{-\infty}^{\infty}
f_F(P)P_\perp^2dP_xdP_y = 2\pi \int\limits_{0}^{\infty}\dfrac{\rho^3d\rho}
{1+e^{\rho^2+\tau^2-\alpha}}=$$$$=2\pi \int\limits_{0}^{\infty}
\dfrac{e^{\alpha-\rho^2-\tau^2}\rho^3d\rho}{1+e^{\alpha-\rho^2-\tau^2}}=
2\pi \int\limits_{0}^{\infty}\rho\ln(1+e^{\alpha-\rho^2})d\rho=2 \pi
f_3(\tau,\alpha).
$$

Здесь
$$
f_3(\tau,\alpha)=\int\limits_{0}^{\infty}x^3f_F(x,\tau,\alpha)dx,\qquad
f_F(x,\tau,\alpha)=\dfrac{1}{1+e^{x^2+\tau^2-\alpha}}.
$$

Согласно (3.9) и (3.10) получаем:
$$
\dfrac{\sigma_{tr}}{\sigma_0}=\dfrac{i \nu}{\omega}\Big[1+
\dfrac{1}{4f_2(\alpha)}\dfrac{\omega I_\omega+i \nu I_\nu}{\omega+i \nu}\Big]
\eqno{(3.11)}
$$
и
$$
\varepsilon_{tr}=1-\dfrac{\omega_p^2}{\omega^2}\Big[1+
\dfrac{1}{4f_2(\alpha)}\dfrac{\omega I_\omega\omega+i \nu I_\nu}
{\omega+i \nu}\Big]
\eqno{(3.12)}
$$

Здесь
$$
I_\nu=\dfrac{1}{2\pi}\int \dfrac{f_F(P)P_\perp^2d^3P}{P_x^2-(l/2)^2}=
\int\limits_{-\infty}^{\infty}\dfrac{f_3(\tau,\alpha)d\tau}{\tau^2-(l/2)^2},
$$
$$
I_\omega=\dfrac{1}{2\pi}\int \dfrac{f_F(P)P_\perp^2d^3P}{(P_x-z/l)^2-(l/2)^2}=
\int\limits_{-\infty}^{\infty}\dfrac{f_3(\tau,\alpha)d\tau}
{(\tau-z/l)^2-(l/2)^2},
$$

Формулу (3.11) и (3.12) перепишем в безразмерных параметрах $x,y,\mathbf{l}$:
$$
\dfrac{\sigma_{tr}}{\sigma_0}=\dfrac{iy}{x}\Big[1+\dfrac{xI_\omega+iyI_\nu}
{4f_2(\alpha)(x+iy)}\Big]
\eqno{(3.13)}
$$
и $$
\varepsilon_{tr}=1-\dfrac{x_p^2}{x^2}\Big[1+
\dfrac{xI_\omega+i y I_\nu}{4f_2(\alpha)(x+iy)}\Big].
\eqno{(3.14)}
$$
Здесь $x_p=\dfrac{\omega_p}{k_Tv_T}$ -- безразмерная плазменная
(ленгмюровская) частота.

Формулы (3.13) и (3.14) будут использованы ниже для проведения численных
и графических расчетов.

\begin{center}
\bf 4. Правило суммы
\end{center}

Проверим выполнение одного из соотношений, называемого правилом $f$--сумм
(см., например, \cite{Dressel}, \cite{Pains} и \cite{Martin}) для поперечной
диэлектрической проницаемости (3.7). Это правило выражается формулой (4.200)
из монографии \cite{Pains}:
$$
\int\limits_{-\infty}^{\infty}\varepsilon_{tr}(\mathbf{q},\omega,\nu)\omega
d\omega=\pi \omega_p^2,
\eqno{(4.1)}
$$
где $\omega_p$ -- плазменная (ленгмюровская) частота,
$$
\omega_p=\sqrt{\dfrac{4\pi e^2 N}{m}}.
$$

Как показано в \cite{Pains}, для доказательства соотношения (4.1) достаточно
доказать выполнение предельного соотношения
$$
\varepsilon_{tr}(\mathbf{q},\omega,\nu)=1-\dfrac{\omega_p^2}{\omega^2}+
o\Big(\dfrac{1}{\omega^2}\Big), \qquad \omega\to \infty.
\eqno{(4.2)}
$$

Воспользуемся выражением (3.3) для поперечной
диэлектрической проницаемости
$$
\varepsilon_{tr}(\mathbf{q},\omega,\nu)=1-\dfrac{\omega_p^2}{\omega^2}
\Bigg[1+$$$$+\dfrac{\hbar^2}{8\pi^3 Nm}\int \Xi(\mathbf{k,q})
[f_F(\mathbf{k})-f_F(\mathbf{k-q})]\mathbf{k}_\perp^2
d\mathbf{k}\Bigg].
\eqno{(4.3)}
$$
Из выражения (4.3) видно, что для доказательства (4.2) достаточно доказать, что
$$
\lim\limits_{\omega\to\infty}\Bigg[1+\dfrac{\hbar^2}{8\pi^3 Nm}\int
\Xi(\mathbf{k,q})
[f_F(\mathbf{k})-f_F(\mathbf{k-q})]\mathbf{k}_\perp^2
d\mathbf{k}\Bigg]=1.
$$
Последнее соотношение совершенно очевидно, если заметить, что
$$
\lim\limits_{\omega\to\infty}\Xi(\mathbf{k,q})=0.
$$

Таким образом, правило $f$--сумм \cite{Pains} для поперечной диэлектрической
проницаемости квантовой столновительной плазмы выполняется.

\begin{center}
\bf 5. Частные случаи электрической проводимости
\end{center}

Покажем, что
$$
\lim\limits_{\omega\to 0}\sigma_{tr}=\sigma_0.
\eqno{(5.1)}
$$
Возьмем формулу (3.7). Заметим, что третье слагаемое в (3.7) пропорционально
$\omega$. Далее, легко видеть, что
$$
f_F(\mathbf{k})=\dfrac{1}{1+e^{P^2}},\quad
f_F(\mathbf{k-q})=\dfrac{1}{1+e^{(P_x-q/k_T)^2+P_\perp^2}}, \quad
P_\perp^2=P_y^2+P_z^2.
$$
При малых $l=q/k_T$ из последней формулы получаем:
$$
f_F(\mathbf{k-q})=f_F(\mathbf{k})+g(P)2P_x\dfrac{q}{k_T},\qquad
g(P)=\dfrac{e^{P^2-\alpha}}{(1+e^{P^2-\alpha})^2}.
$$

Согласно (3.7) находим, что
$$
\dfrac{\sigma_{tr}}{\sigma_0}=\dfrac{i \nu}{\omega}\Bigg[1-
\dfrac{i \nu}{4\pi f_2(\alpha)(\omega+i \nu)}\int
\dfrac{g(P)P_xP_\perp^2d^3P}{P_x-q/2k_T}\Bigg].
$$
Устремим $q\to 0$ в этом выражении. Получаем, что
$$
\dfrac{\sigma_{tr}}{\sigma_0}=\dfrac{i \nu}{\omega}\Bigg[1-
\dfrac{i \nu}{4\pi f_2(\alpha)(\omega+i \nu)}\int g(P)P_\perp^2d^3P\Bigg].
\eqno{(5.2)}
$$
Рассмотрим интеграл из (5.2):
$$
\int g(P)P_\perp^2d^3P=2\pi \int\limits_{-\infty}^{\infty}\int\limits_{0}^{\infty}
g(P)P_\perp^3 d^3P=
$$$$=2\pi\int\limits_{0}^{\infty}\ln(1+e^{\alpha-\tau^2})d\tau=
4\pi f_2(\alpha).
$$

На основании (5.2) теперь получаем:
$$
\sigma_{tr}=\sigma_0\dfrac{i \nu}{\omega+i \nu}.
$$
Отсюда видно, что при $\omega\to 0$ и при малых $q$ поперечная проводимость
квантовой плазмы переходит в статическую проводимость классической плазмы.

Пусть теперь величина $\omega$ мала, но не стремится к нулю. Преобразуем
формулу (3.7) к следующему виду:
$$
\dfrac{\sigma_{tr}}{\sigma_0}=\dfrac{i \nu}{\omega}\Bigg[1-
\dfrac{i \nu}{4\pi f_2(\alpha)(\omega+i \nu)}\int \dfrac{g(P)P_xP_\perp^2d^3P}
{P_x-q\hbar/2mv_T}- $$
$$
-\dfrac{\omega}{4\pi f_2(\alpha)(\omega+i \nu)}\int\dfrac{g(P)P_xP_\perp^2d^3P}
{P_x-q\hbar/2mv_T-(\omega+i \nu)/v_Tq}\Bigg].
\eqno{(5.3)}
$$
Вычислим внутренний двойной интеграл по плоскости $(P_y,P_z)$ в формуле (5.3):
$$
\int\limits_{-\infty}^{\infty}\int\limits_{-\infty}^{\infty}
g(P)P_\perp^2dP_ydP_z=\pi\ln(1+e^{\alpha-P_x^2}).
$$

Теперь формула (5.3) упрощается:
$$
\dfrac{\sigma_{tr}}{\sigma_0}=\dfrac{i \nu}{\omega}\Bigg[1-
\dfrac{i \nu}{4f_2(\alpha)(\omega+i \nu)}\int\limits_{-\infty}^{\infty}
\dfrac{P_x\ln(1+e^{\alpha-P_x^2})dP_x}{P_x-l/2}-$$$$-\dfrac{\omega}{4f_2(\alpha)
(\omega+i \nu)}\int\limits_{-\infty}^{\infty}\dfrac{P_x\ln(1+e^{\alpha-P_x^2})}
{P_x-l/2-z/l}\Bigg],\quad l=\dfrac{q}{k_T},\quad z=\dfrac{\omega+i \nu}{k_Tv_T},
$$
или
$$
\dfrac{\sigma_{tr}}{\sigma_0}=\dfrac{i \nu}{\omega}\Bigg[1-
\dfrac{i y}{4f_2(\alpha)z}\int\limits_{-\infty}^{\infty}
\dfrac{\tau\ln(1+e^{\alpha-\tau^2})d\tau}{\tau-l/2}- $$$$-
\dfrac{x}{4f_2(\alpha)
z}\int\limits_{-\infty}^{\infty}\dfrac{\tau\ln(1+e^{\alpha-\tau^2})}
{\tau-l/2-z/l}\Bigg],
\eqno{(5.4)}
$$
$$
l=\dfrac{q}{k_T},\qquad z=\dfrac{\omega+i \nu}{k_Tv_T}.
$$

Возьмем выражение поперечной проводимости в классической плазме:
$$
\dfrac{\sigma_{tr}}{\sigma_0}=\dfrac{1}{4\pi f_2(\alpha)}\int
\dfrac{g(P)P_\perp^2d^3P}{1-i\omega\tau+iql_TP_x},\quad  l_T=v_T\tau.
\eqno{(5.5)}
$$
Формула (5.5) упрощается и принимает вид:
$$
\dfrac{\sigma_{tr}}{\sigma_0}=-\dfrac{iy}{4f_2(\alpha)l}
\int\limits_{-\infty}^{\infty}\dfrac{\ln(1+e^{\alpha-\tau^2})d\tau}{\tau-z/l}.
\eqno{(5.6)}
$$

Покажем, что формула (5.4) приводится при малых $l$ к формуле (5.6) (это
формула (3.7) из нашей работы
\cite{Arxiv10} для поперечной электрической проводимости классической плазмы).

В формуле (5.4) преобразуем первый интеграл, разбивая его на два слагаемых.
В результате получаем, что формула (5.4) преобразуется к виду:
$$
\dfrac{\sigma_{tr}}{\sigma_0}=\dfrac{i \nu}{\omega}\Big[\dfrac{\omega}
{\omega+i \nu}-\dfrac{i \nu}{\omega+i \nu}\dfrac{1}{4f_2(\alpha)}
\dfrac{l}{2}\int\limits_{-\infty}^{\infty}\dfrac{\ln(1+e^{\alpha-\tau^2})d\tau}
{\tau-l/2}-$$$$-
\dfrac{\omega}{\omega+i \nu}\dfrac{1}{4f_2(\alpha)}
\int\limits_{-\infty}^{\infty}\dfrac{\tau\ln(1+e^{\alpha-\tau^2})d\tau}
{\tau-l/2-z/l}\Big].
\eqno{(5.7)}
$$

Из (5.7) видно, что первый интеграл пропорционален $l^2$. Отбросим этот интеграл.
В знаменателе второго интеграла из (5.7) пренебрегаем членом $l/2$, ибо
$l\ll |z|/l$. В результате для малых значений $l$ получаем:
$$
\dfrac{\sigma_{tr}}{\sigma_0}=\dfrac{i \nu}{\omega+i \nu}\Big[1-
\dfrac{1}{4f_2(\alpha)}\int\limits_{-\infty}^{\infty}
\dfrac{\tau\ln(1+e^{\alpha-\tau^2})}{\tau-z/l}d\tau\Big].
\eqno{(5.8)}
$$
Преобразуя интеграл из (5.8) на два, мы в точности получаем формулу
(5.6):
$$
\dfrac{\sigma_{tr}}{\sigma_0}=-\dfrac{iy}{l4f_2(\alpha)}
\int\limits_{-\infty}^{\infty}\dfrac{\ln(1+e^{\alpha-\tau^2})}
{\tau-z/l}d \tau.
$$

На рис. 1 -- 6 представим сравнение модулей электрических проводимостей
классической (кривые "1") и квантовой плазмы (кривые "2") в зависимости от
 безразмерной частоты колебаний электрического поля.

На рис. 7 и 8 представлены зависимости действительной и мнимой частей
электрической проводимости квантовой плазмы от безразмерного волнового
числа (рис. 7) и от безразмерной частоты колебаний
электрического поля (рис. 8).

На рис. 9 и 10  представлены зависимости действительной части (рис. 9)
и мнимой части (рис. 10) электрической проводимости от безразмерной частоты
колебаний электрического поля.
\begin{center}
\bf 6. Заключение
\end{center}

В настоящей работе выведена формула для электрической
проводимости в квантовой невырожденной столкновительной плазме с произвольной
степенью вырожденности электронного газа.
Для этой цели используется кинетическое уравнение с интегралом столкновений
в форме релаксационной модели в пространстве импульсов.
Проводится графическое сравнение квантовой проводимости из настоящей работы с
классической проводимостью.
\begin{figure}[h]
\begin{center}
\includegraphics[width=17.0cm, height=10cm]{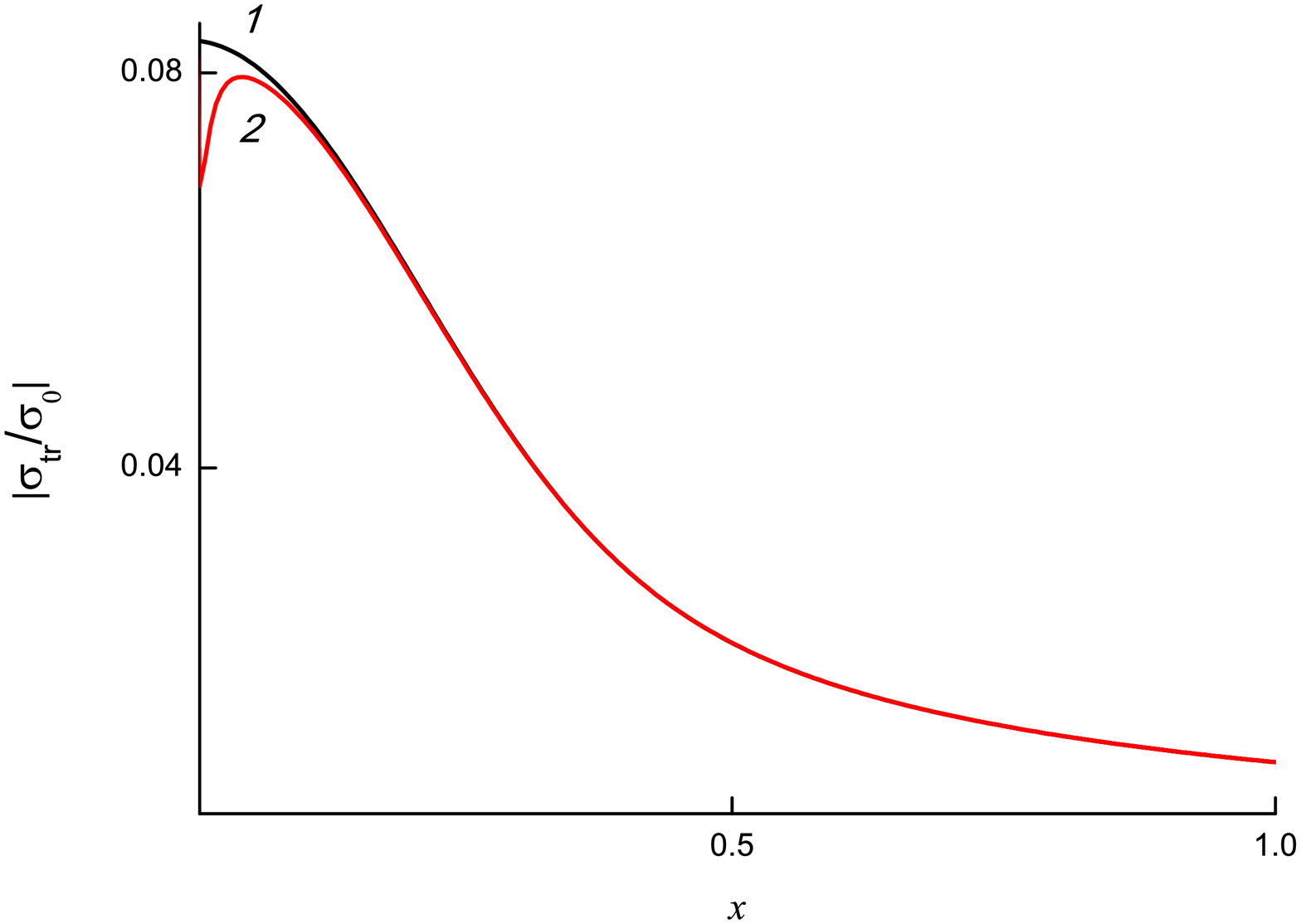}
\end{center}
\begin{center}
{{ Fig. 1. Dependence of $|\sigma_{tr}/\sigma_0|$ on quantity $x$; $y=0.01,
q=0.2$, $\alpha=-3$.}}
\end{center}
\end{figure}

\begin{figure}[h]
\begin{center}
\includegraphics[width=17.0cm, height=10cm]{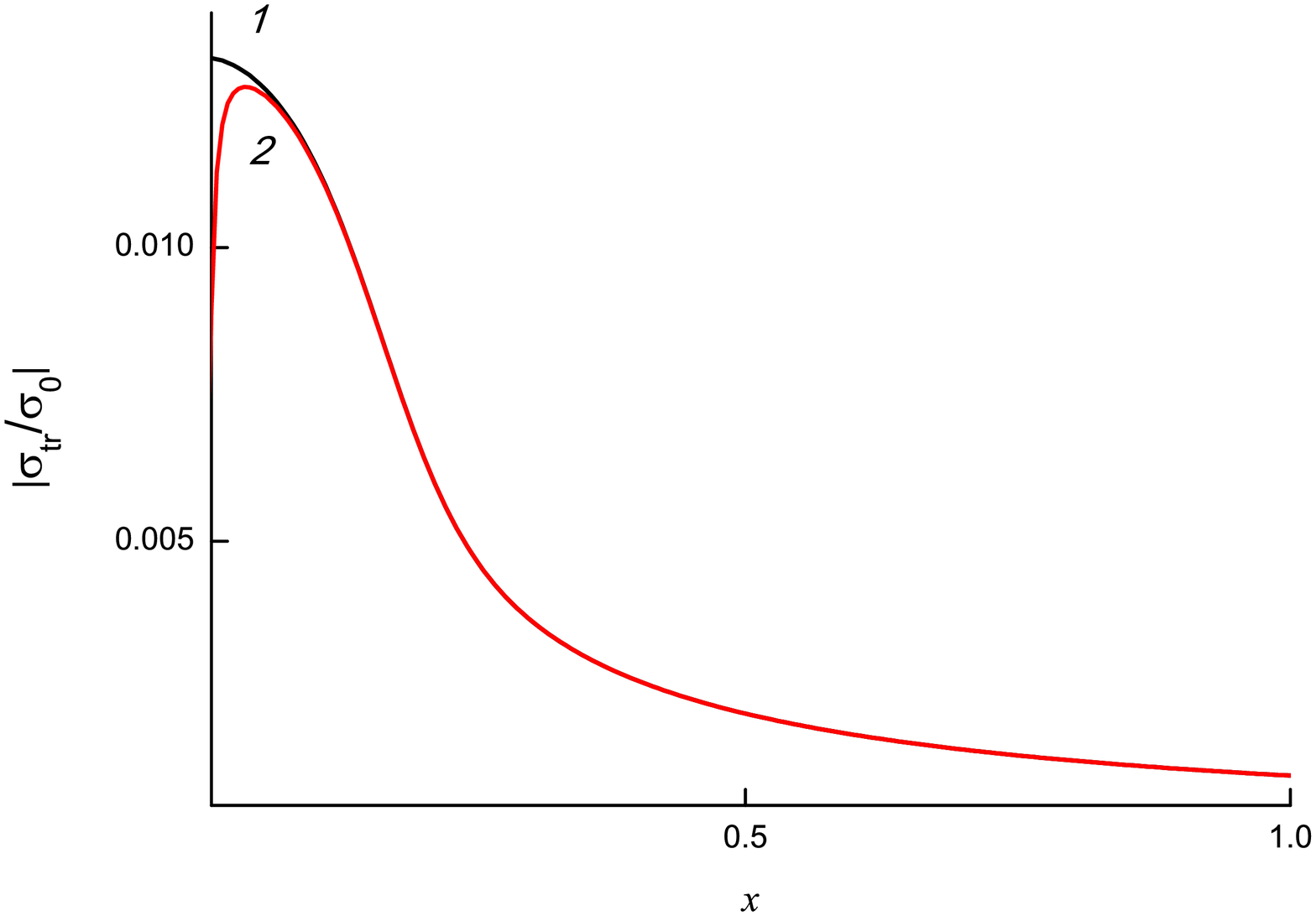}
\end{center}
\begin{center}
{{ Fig. 2. Dependence $|\sigma_{tr}/\sigma_0|$ of quantity $x$; $y=0.001, q=0.1$,
$\alpha=2$.}}
\end{center}
\end{figure}

\begin{figure}[h]
\begin{center}
\includegraphics[width=17.0cm, height=10cm]{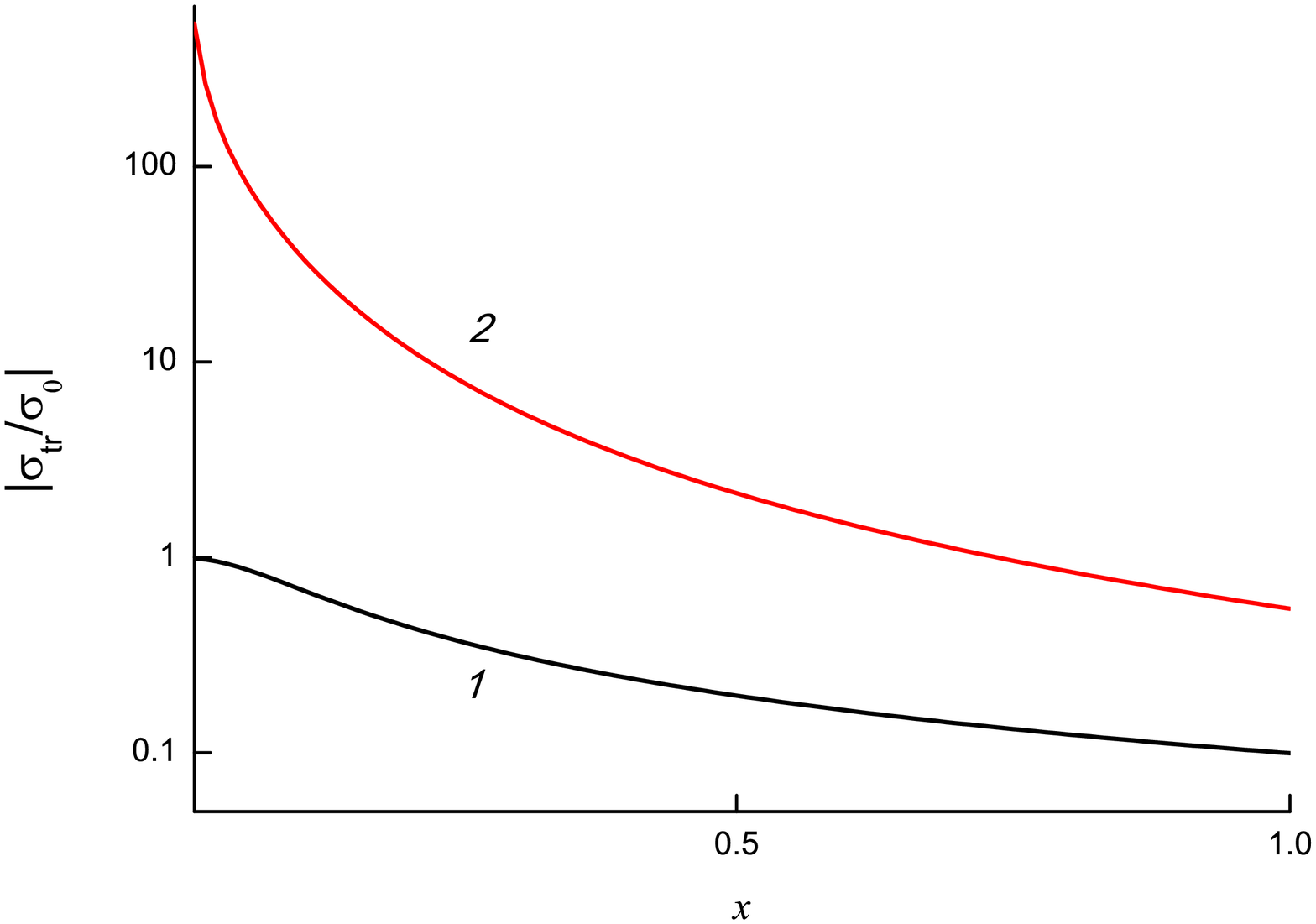}
\end{center}
\begin{center}
{{ Fig. 3. Dependence of $|\sigma_{tr}/\sigma_0|$ on quantity $x$; $y=0.1,
q=0.01$, $\alpha=-3$.}}
\end{center}
\end{figure}

\begin{figure}[h]
\begin{center}
\includegraphics[width=17.0cm, height=10cm]{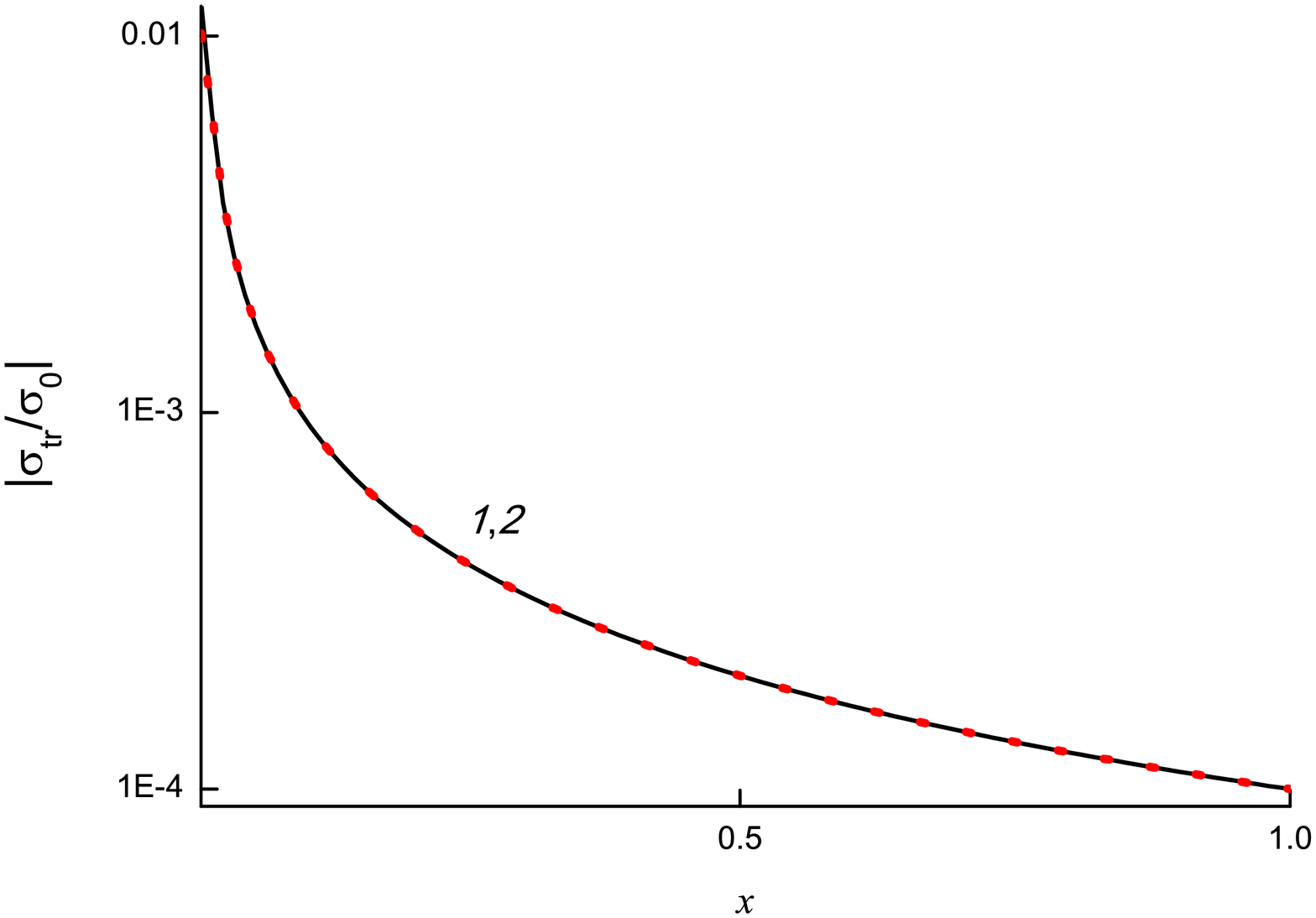}
\end{center}
\begin{center}
{{ Fig. 4. Dependence of $|\sigma_{tr}/\sigma_0|$ on quantity $x$; $y=0.0001,
q=0.01$, $\alpha=0$.}}
\end{center}
\end{figure}

\begin{figure}[h]
\begin{center}
\includegraphics[width=17.0cm, height=10cm]{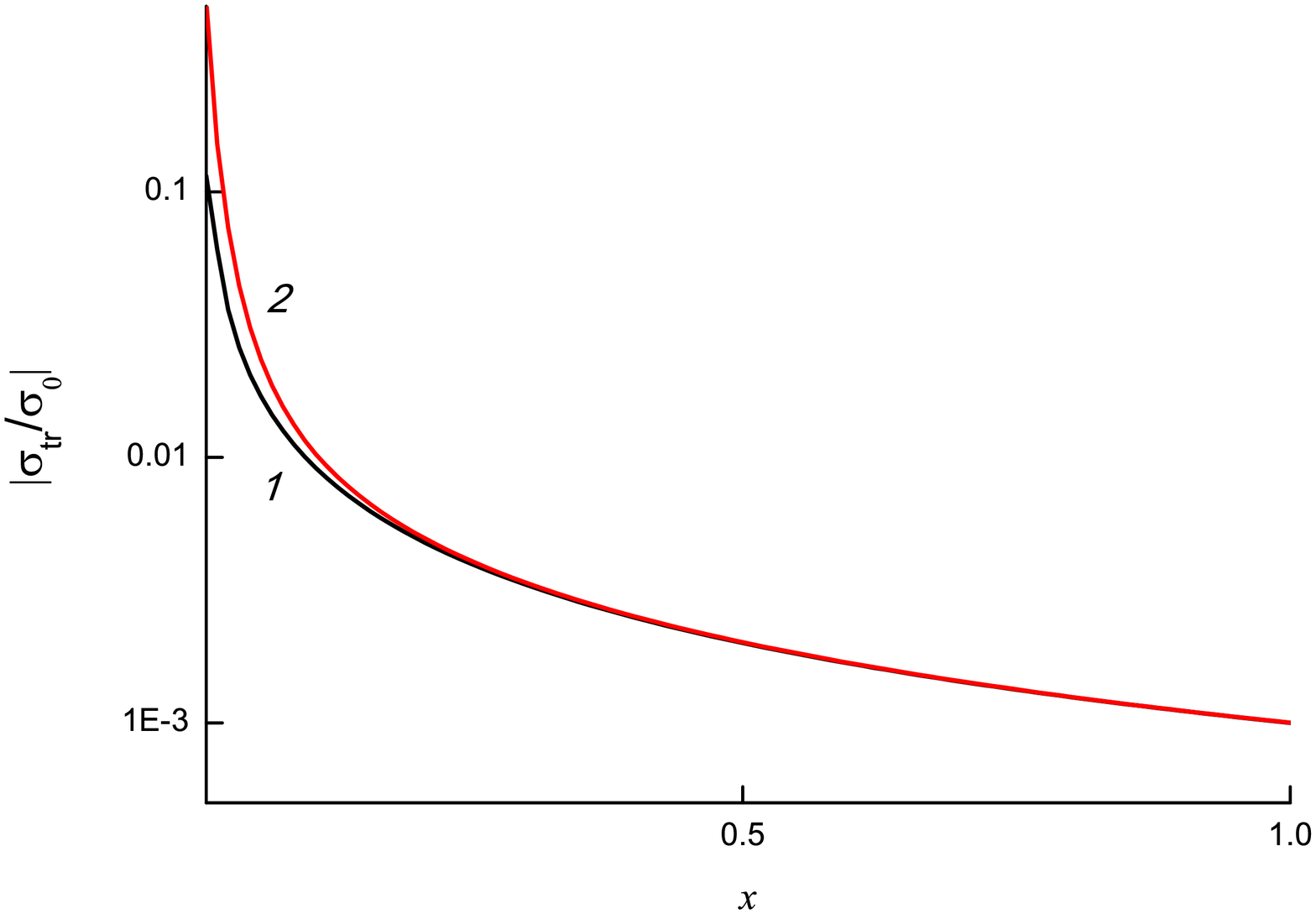}
\end{center}
\begin{center}
{{ Fig. 5. Dependence of $|\sigma_{tr}/\sigma_0|$ on quantity $x$;
$y=0.001, q=0.1, \alpha=0$.}}
\end{center}
\end{figure}

\begin{figure}[h]
\begin{center}
\includegraphics[width=17.0cm, height=10cm]{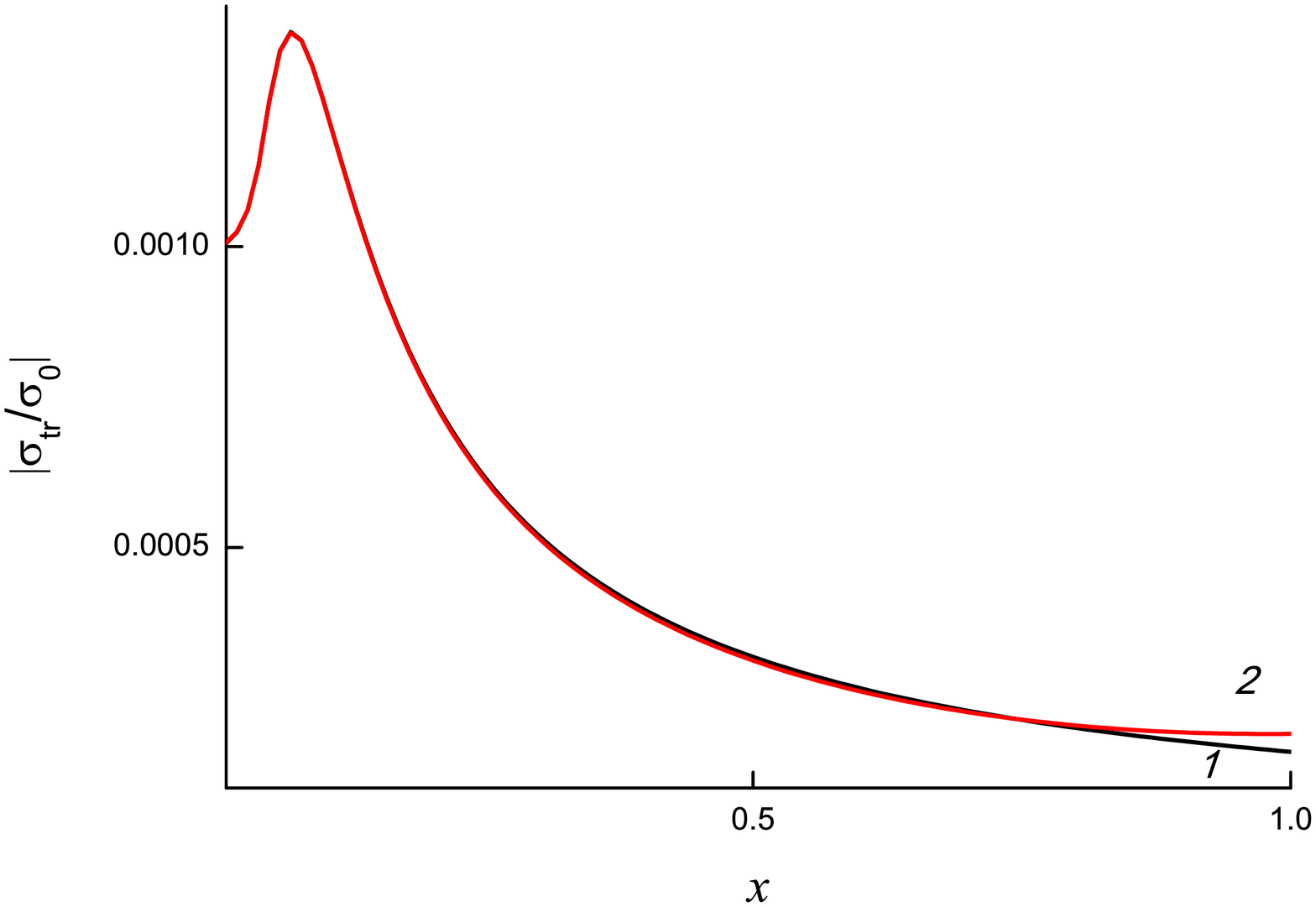}
\end{center}
\begin{center}
{{ Fig. 6. Dependence of $|\sigma_{tr}/\sigma_0|$ on quantity $q$;
$x=0.1, y=0.0001, 0\leqslant q \leqslant 1$, $\alpha=0$.}}
\end{center}
\end{figure}

\begin{figure}[h]
\begin{center}
\includegraphics[width=17.0cm, height=10cm]{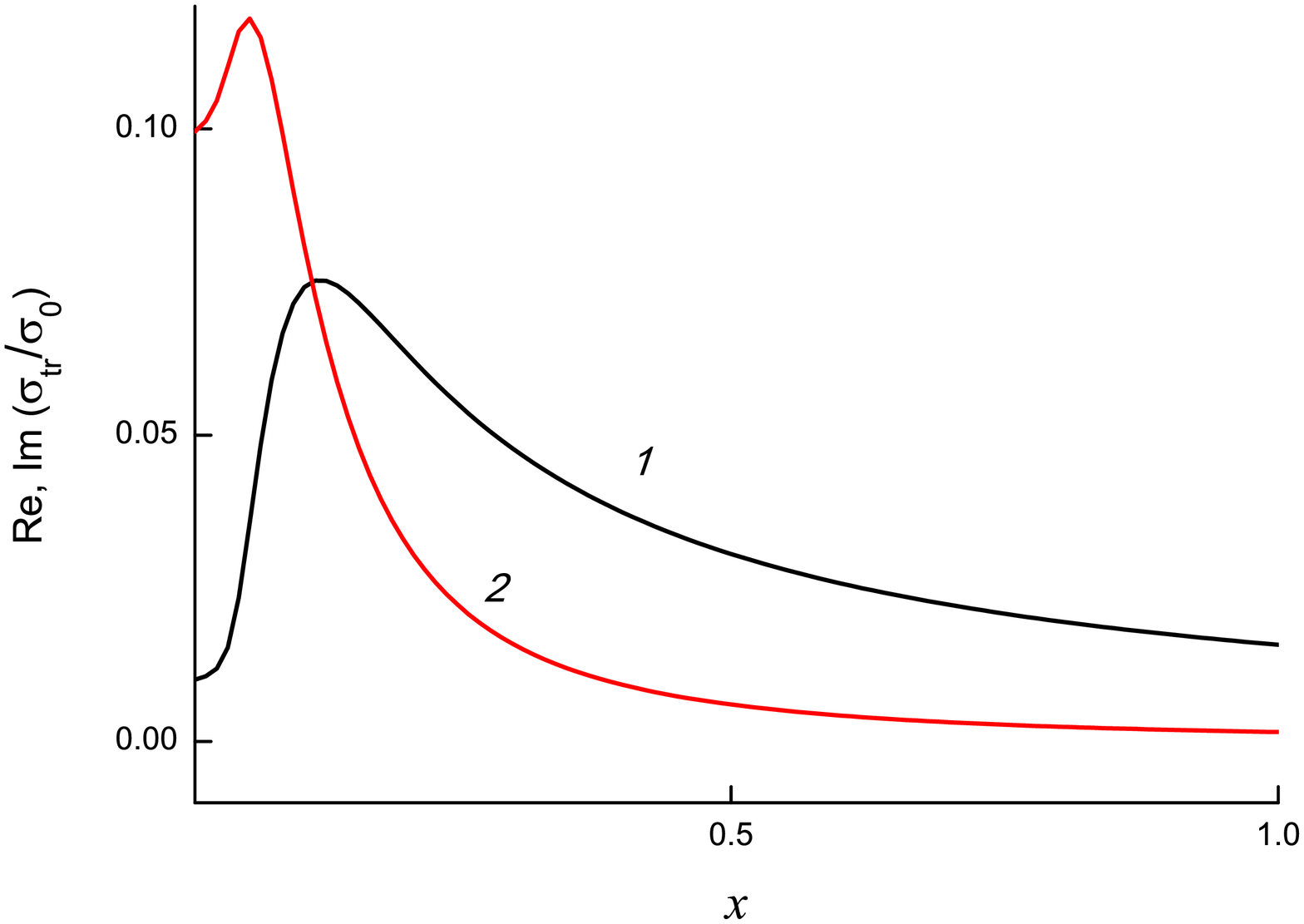}
\end{center}
\begin{center}
{{ Fig. 7. Dependence of $\Re(\sigma_{tr}/\sigma_0)$ (curve "1") and
$\Im(\sigma_{tr}/\sigma_0)$ (curve "2") on quantity $q$;
$x=0.1, y=0.01$, $\alpha=0$.}}
\end{center}
\end{figure}

\begin{figure}[h]
\begin{center}
\includegraphics[width=17.0cm, height=10cm]{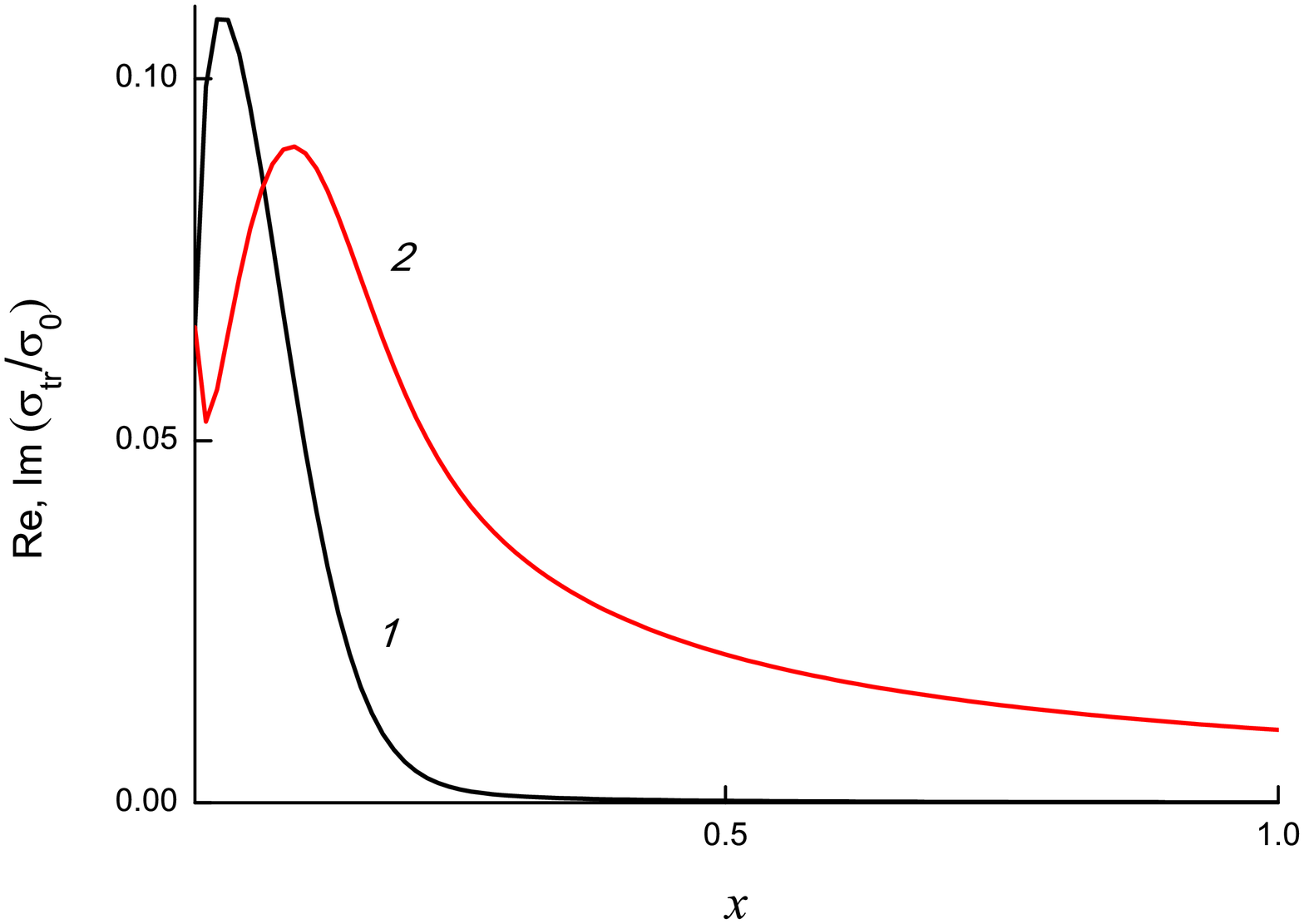}
\end{center}
\begin{center}
{{ Fig. 8. Dependence of $\Re(\sigma_{tr}/\sigma_0)$ (curve "1") and
$\Im(\sigma_{tr}/\sigma_0)$ (curve "2") on quantity $x$;
$y=0.01$, $q=0.1$, $\alpha=0$.}}
\end{center}
\end{figure}

\begin{figure}[h]
\begin{center}
\includegraphics[width=17.0cm, height=10cm]{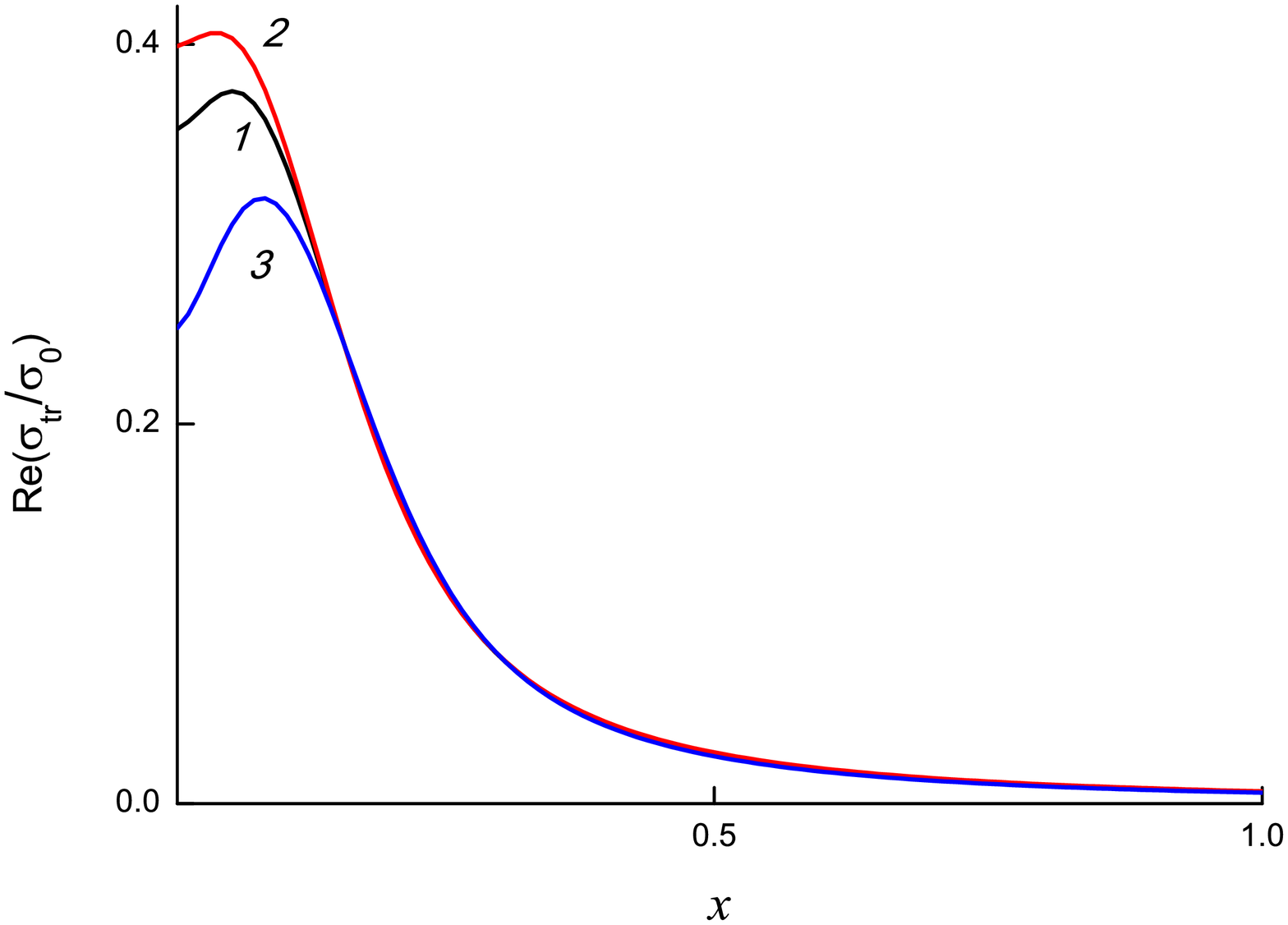}
\end{center}
\begin{center}
{{ Fig. 9. Dependence of $\Re(\sigma_{tr}/\sigma_0)$ on quantity $x$;
$y=0.1$, $q=0.1$; сurves of $1,2,3$ correspond to values of parameter $\alpha$:
$\alpha=0,-3,2$.}}
\end{center}
\end{figure}

\begin{figure}[h]
\begin{center}
\includegraphics[width=17.0cm, height=12cm]{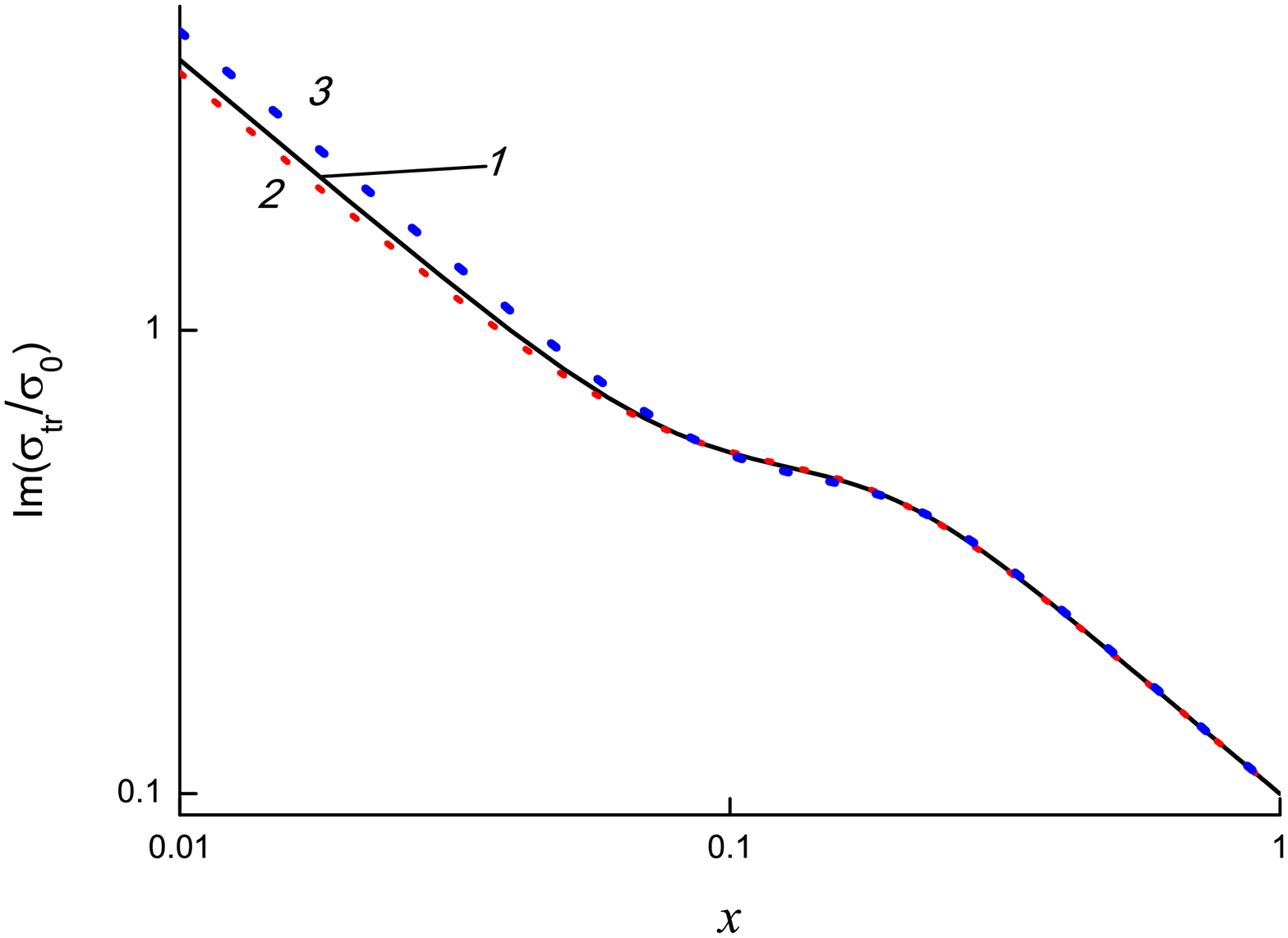}
\end{center}
\begin{center}
{{ Fig. 10. Dependence of $\Im(\sigma_{tr}/\sigma_0)$ on quantity $x$;
$y=0.1$, $q=0.1$; curves $1,2,3$ correspond to values of parameter
$\alpha$: $\alpha=0, -3,2$.}}
\end{center}
\end{figure}

\end{document}